\documentclass[apj]{emulateapj}
\usepackage{natbib}
\usepackage{color}
\usepackage{epsfig}
\usepackage{amsmath}
\usepackage{natbib}
\usepackage{bm}

\newcommand{\um}{$\, \mu$m }
\newcommand{\mll}{$M_{\ell\ell'}$ }

\newcommand{\m}{$\mu$m}
\newcommand{\ch}{{\it Chandra }}
\newcommand{\cxb}{CIB $\times$ CXB }

\begin{document}

\title{Cross-correlation between X-ray and optical/near-infrared background intensity fluctuations}
\author{Ketron Mitchell-Wynne$^{1}$, Asantha Cooray$^{1}$, Yongquan Xue$^{2}$, Bin Luo$^{3,4}$, William Brandt$^{5,6,7}$, Anton Koekemoer$^{8}$}
\affiliation{$^{1}$ Department of Physics \& Astronomy, University of California, Irvine, CA 92697}
\affiliation{$^{2}$ CAS Key Laboratory for Researches in Galaxies and Cosmology, Center for Astrophysics, Department of Astronomy, University of Science and Technology of China, Chinese Academy of Sciences, Hefei, Anhui 230026, China}
\affiliation{$^3$ School of Astronomy and Space Science, Nanjing University, Nanjing, 210093, China}
\affiliation{$^4$ Key laboratory of Modern Astronomy and Astrophysics (Nanjing University), Ministry of Education, Nanjing 210093, China}
\affiliation{$^{5}$ Department of Astronomy \& Astrophysics, Pennsylvania State University, University Park, PA, 16802}
\affiliation{$^{6}$ Institute for Gravitation and the Cosmos, Pennsylvania State University, 525 Davey Lab, University Park, PA 16802}
\affiliation{$^{7}$  Department of Physics, 104 Davey Lab, The Pennsylvania State University, University Park, PA 16802, USA}
\affiliation{$^{8}$ Space Telescope Science Institute, 3700 San Martin Drive, Baltimore, MD 21218, USA}

\begin{abstract}
Angular power spectra of optical and infrared background anisotropies at wavelengths between 0.5 to 5 $\mu$m are a useful probe of faint sources present during reionization, in addition to faint galaxies and diffuse signals 
at low redshift. The cross-correlation of these fluctuations with backgrounds at other wavelengths can be used to separate some of these signals. A previous study on the cross-correlation between \mbox{X-ray} and {\it Spitzer} fluctuations at 3.6 $\mu$m and 4.5 $\mu$m has been interpreted as evidence for direct collapse blackholes (DCBHs) present at $z > 12$. Here we return to this cross-correlation and study its wavelength dependence from 0.5 to 4.5 $\mu$m using {\it Hubble} and {\it Spitzer} data in combination with a subset of the 4 Ms {\it Chandra} observations in GOODS-S/ECDFS. Our study involves five {\it Hubble} bands at 0.6, 0.7, 0.85, 1.25 and 1.6~$\mu$m, and two {\it Spitzer}-IRAC bands at 3.6~$\mu$m and 4.5~$\mu$m. We confirm the previously seen cross-correlation between 3.6 $\mu$m (4.5~$\mu$m) and \mbox{X-rays} with 3.7$\sigma$ (4.2$\sigma$) and 2.7$\sigma$ (3.7$\sigma$) detections in the soft [0.5-2] keV and hard [2-8] keV X-ray bands, respectively, at angular scales above 20 arcseconds. The cross-correlation of X-rays with {\it Hubble} is largely anticorrelated, ranging between the levels of 1.4$-$3.5$\sigma$ for all the {\it Hubble} and X-ray bands. This lack of correlation in the shorter optical/NIR bands implies the sources responsible for the cosmic infrared background at 3.6 and 4.5~$\mu$m are at least partly dissimilar to those at 1.6 $\mu$m and shorter. 
\end{abstract}

\section{Introduction}

In addition to galaxies throughout cosmic history, the cosmic infrared background (CIB) also contains the signatures from diffuse sources of emission \citep{Santos02}, such as the intra-halo light (IHL; \citealt{Cooray12, Zemcov14}), or sources that are not traditionally counted as galaxies. For example, the near-IR background (NIRB) is generally considered to be a probe of reionization and the sources present during that epoch. Indeed it was first posited by \cite{Kashlinsky05} that NIRB clustering measured on arcmin scales was due to zero-metallicity Population III stars during reionization. Theoretical calculations by \cite{Cooray12b} subsequently showed that the measured amplitude in previous works claiming Population III signatures \citep{Kashlinsky07, Matsumoto11} is an order of magnitude larger than theoretical expectations \citep{Komatsu11}. This shifted the discussion to include the possibility of a number of other sources contributing to the NIRB, including low-redshift galaxies \citep{Sullivan07, Thompson07, Cooray07, Chary08}. The contribution from low-$z$ galaxies was then explicitly constrained by \cite{Helgason12} by reconstructing the background signal from the known low-$z$ galaxy populations.

There still exists an excess signal in the NIRB fluctuation power spectra, in addition to the Pop III signature and low-$z$ galaxy population. As mentioned in the beginning, one such low redshift contribution is now identified to be IHL \citep{Cooray12, Zemcov14}. An additional and important source population to consider involves direct collapse black holes (DCBHs; e.g. \citealt{Yue13, Yue14}) that are hypothesized to explain the presence of massive billion to ten billion solar mass black holes in luminous quasars at $z \sim 6$ (e.g. \citealt{Fan06}). Actively accreting DCBHs are expected to leave an imprint in the X-ray background. Similarly, the rest-frame UV photons emitted by these sources will now be visible in the infrared. If these sources were present during the earliest epochs of galaxy and black hole formation, it would then make sense that the cosmic X-ray background (CXB) would be correlated with the intensity fluctuations of the near-IR background.

This cross-correlation was measured to be statistically significant at 3.6 and 4.5~\m~ in the {\it Spitzer}/Extended Groth Strip (EGS; \citealt{Davis07}) with the broad-band 0.5-2 keV background fluctuations (\citealt{Cappelluti13}, hereafter C13) of the same field. Cross-correlations between {\it Spitzer} and the harder X-ray band at 2-7 keV were found to be statistically less significant. This detection in the soft band and the non-detection in the hard band was subsequently modeled in \cite{Helgason14} where they found that X-ray emission from AGN, normal galaxies and virialized hot gas at $z\,\lesssim\,6$ cannot account for the large-scale correlation. They suggest that the CIB and CXB signals are from the same, unknown source population at high-$z$, perhaps miniquasars as proposed in \cite{Cappelluti12} and explained as DCBHs in \cite{Yue13b}. In order to gain more insight, and to test whether the cross-correlation is present in other legacy fields, it is essential that we consider more optical and infrared bands and extend the studies to other areas.

The Cosmic Assembly Near-infrared Deep Extragalactic Legacy Survey (CANDELS; \citealt{Grogin11, Koekemoer11}) has opened up a new window of opportunity in measuring the intensity fluctuations at near-infrared (NIR) and optical wavelengths, ranging between 0.6 and 1.6~\m. Statistical measurements (namely the angular power spectrum) of the background fluctuations at this wavelength range were only recently measured (\citealt{Mitchell-Wynne15}, hereafter MW15) and are interpreted to have contributions from IHL, faint low-$z$ galaxies \citep{Helgason12}, possibly diffuse Galactic light (DGL) from dust in our Milky Way, and first-light galaxies during the reionization epoch in the two 1.25 and 1.6 $\mu$m {\it Hubble}/WFC3 bands. Angular power spectra measurements have also been made over a 10 deg$^{2}$ field in the IRAC 3.6 and 4.5~\m~ bands, with similar interpretations of IHL and faint low-$z$ residual galaxies \citep{Cooray12}, with no contribution from high-redshift sources. The measurements with IRAC by independent groups (e.g., Kashlinsky et al. 2007; 2012) agree with each other, though interpretations of those clustered fluctuations remain different.

In this paper we show results of \cxb cross-correlations in the \ch Deep Field-South (CDF-S/GOODS-S; \citealt{Giavalisco04}), using data from one of the deepest \ch X-ray surveys, deep {\it Hubble} exposures from CANDELS and archival HST data, and a combination of a large collection of surveys in IRAC channels~1 and 2. In Section~\ref{sec:maps} we describe our initial data reduction and map-making techinques, and in Section~\ref{sec:ps} we present our cross-correlation measurements, which are discussed in context in Section~\ref{sec:discussion}.

\section{Map making}
\label{sec:maps}
\begin{figure}
  \includegraphics[scale=.66]{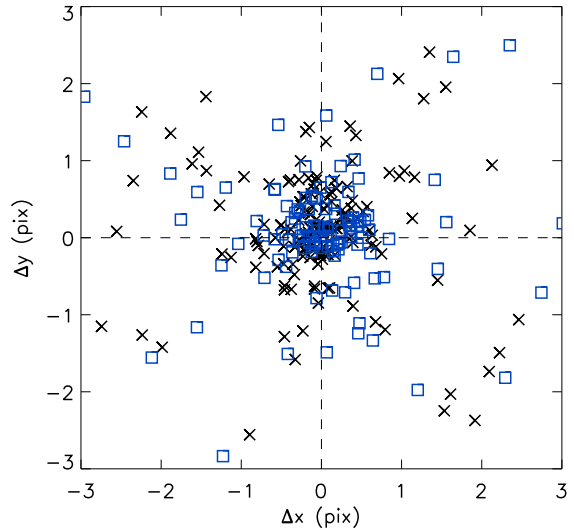}
  \caption{Astrometric alignment between the X-ray and optical/NIR maps after performing the astrometric shift with \texttt{tweakreg}. IRAC detected sources matched with X-ray are denoted as black crosses, and HST sources matched with X-ray are blue squares. Prior to the astrometric shift with \texttt{tweakreg}, the offset was $\sim$~0.1$-$1 pixel at the IRAC pixel scale.}
  \label{fig:align}
\end{figure}

We have assembled maps from three different space telescopes to perform the cross-correlations. For the power spectrum analyses we perform jack-knife tests, so we generate two maps for each filter with the exact same sky coverage by subdiving the exposure time into two subsets. Each of the IRAC and \ch maps have large area coverage, with a limiting area defined by the CANDELS observations; both the CANDELS and \ch maps have small pixel scales with 0.$''$14 and 0.$''$5 pix$^{-1}$ respectively. Thus for all the aligned maps which are used in this work, we are limited by the native IRAC pixel scale at 1.2$''$~pix$^{-1}$, and a $\sim$~110 arcmin$^2$ survey area, defined by the CANDELS observations. Subsequent to generating mosaics for each instrument, they are regridded and cropped to these specifications. Since the mosaics from different instruments are each using different astrometric references, the astrometric alignment between the three was slightly shifted, by 0.1 to 1 pixel. We removed this offset by aligning the X-ray and {\it Spitzer} maps to the HST astrometry using the \texttt{tweakreg} PyRAF module (see Fig~\ref{fig:align}). 

\subsection{{\it Chandra} X-ray maps}
We used \ch data of ECDF-S from Cycle 8 to Cycle 12 as listed in Table~\ref{tab:obs}. The data span 4~Ms of observations and we make two maps with an effective exposure time of $\sim$~1.5~Ms each, generated from a total of 43 observations. The first mosaic is generated from 19 observations, and the second from 24 observations. The details of these observations are listed in Table~\ref{tab:obs}, which are a subset of those listed in Table~1 of \cite{Luo08} and Table~1 of \cite{Xue11}. All observations were done in VFAINT mode. The collection of \ch frames we use here represents the VFAINT subset of the total 4~Ms observations, as it omits all observations done in FAINT mode. Basic processing of the data was done using the \ch X-ray Center (CXC) pipeline. To ensure uniformity, mosaics were produced with the \ch Interactive Analysis of Observations (\texttt{CIAO}), following the details described in \cite{Giacconi02}, \cite{Alexander03} and \cite{Xue11}. However for clarity, we will briefly review the data reduction and analysis procedure here.

Background light curves were inspected using \texttt{EVENT\_BROWSER} in the Tools for ACIS Real-time Analysis (TARA; \citealt{Broos00}). All data were corrected for the radiation damage of the CCDs during the first few months of {\it Chandra} operations using the charge transfer inefficienty correction procedure of \cite{Townsley00, Townsley02}. All bad columns, bad pixels and cosmic ray afterglows were removed using the ``status'' information in the event files, and only observations taken within the CXC-generated good-time intervals was used. \texttt{ACIS\_PROCESS\_EVENTS} was used to identify potential background events and to remove the standard pixel randomization. 

The observations are then split into three standard broad bands: \mbox{0.5-2}~keV (``soft''), \mbox{2-8}~keV (``hard'') and \mbox{0.5-8}~keV (``full''). The mosaics cover a total area of $\sim$~450 arcmin$^2$ (before alignment to the CANDELS field of view), and reach an on-axis sensitivity of 5.1 $\times\,10^{-18}$, 3.7 $\times\,10^{-17}$ and 2.4 $\times\,10^{-17}$ erg s$^{-1}$ cm$^{-2}$ respectively for the soft, hard and full bands \citep{Lehmer12}. Our final merged event file includes only 0.3-10 keV events from which make our mosaics.

The background in the X-ray mosaics contains a high energy particle component in addition to the cosmic signal in which we are interested. The spectral shape of the particle background is fairly flat in the soft and hard bands, and above $\sim$9 keV is solely a product of non-astrophysical particle events. Per the prescription in \cite{Hickox06}, we can isolate the CXB in our X-ray maps by subtracting off the particle background measured in the 9-10 keV exposures. The particle background in the ACIS detectors is a product of interactions of the CCDs and surrounding materials with high-energy particles. Taking ACIS exposures while the instrument is stowed within the detector housing blocks celestial X-rays and thus accurately represents the particle background of the detector.

A stowed image is produced from CALDB version 4.7. A particle background map is then approximated by simply normalizing the stowed image as $C_{\mathrm{data}}[9$-$10] / C_{\mathrm{stow}}[9$-$10]$, as is discussed in \cite{Hickox06}, where the counts for $C_{\mathrm{data}}$ and $C_{\mathrm{stow}}$ are respectively measured from the total counts in the real images and stowed images. We then subtract this particle background from each of our pointings, the same as was done in C13. This will effectively remove the particle background from our analysis, and is an additional step to the reduction procedures detailed in \cite{Xue11}. We exclude energies above 10 keV because all events above that threshold were filtered out in the initial reduction.

In \cite{Hickox06}, they state that data collected in VFAINT mode are the most reliable for background sensitive measurements, however their two other datasets in FAINT mode provide consistent results. In C13 they only use VFAINT data; for the sake of consistency with C13, we have omitted all FAINT observations, which account for $\sim$~1~Ms of exposure time in CDF-S. In addition to the results presented here, we also performed our analysis with the full 4~Ms data set and obtained very similar results, which is consistent with the results of \cite{Hickox06}.

\begin{figure}
  \includegraphics[scale=.4]{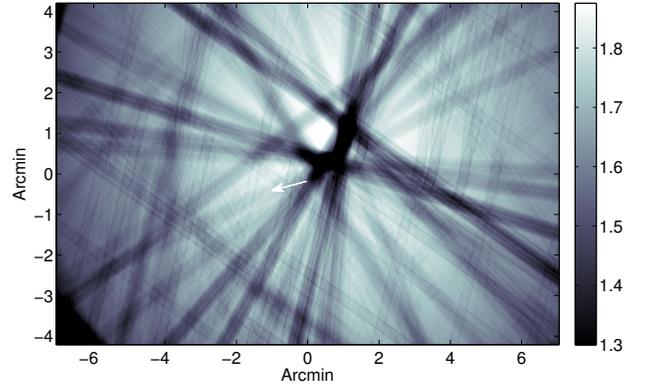}
  \caption{Effective exposure time map for the soft band, in the first half CDF-S mosaic, in units of Ms. The white arrow points North from the center of the map.}
    \label{fig:exposure}
\end{figure}

We convert \ch event maps into flux maps in units of erg s$^{-1}$ cm$^{-2}$ by dividing by the effective exposure time maps (see Fig.~\ref{fig:exposure}), and assuming a power-law index of 1.4 and a Galactic neutral hydrogen column density of 8.8~$\times\,10^{19}$~cm$^{-2}$. With these criteria, we used the NASA's HEASARC online tool (WebPIMMS; https://heasarc.gsfc.nasa.gov/cgi-bin/Tools/w3pimms/w3pimms.pl) to obtain conversion factors (ecf) for the soft, hard, and full bands of 2.15~$\times\,10^{10}$, 1.90~$\times\,10^{10}$ and 3.53~$\times\,10^{10}$~erg$^{-1}$~cm$^{2}$, respectively. These mosaics are then aligned to the CANDELS field of view, repixelized to 1.2$''$ pix$^{-1}$, and smoothed by a 3.6$''$ Gaussian kernel. The full mosaic, along with the HST CDF-S footprint can be seen in Figure~1 of \cite{Xue11}.

\begin{deluxetable*}{lcccccl}
\tabletypesize{\small}
\tablecaption{Journal of {\it Chandra} Deep Field-South Observations}
\tablehead{
\colhead{}                                 &
\colhead{Obs. Start}                                 &
\colhead{Exposure}                             &
\multicolumn{2}{c}{Aim Point}                 &
\colhead{Roll Angle}                                 &
\colhead{}                                \\
\cline{4-5}
\colhead{Obs. ID}                                 &
\colhead{(UT)}                         &
\colhead{Time (ks)}               &
\colhead{$\alpha$ (J2000.0)}                &
\colhead{$\delta$ (J2000.0)}                &
\colhead{(deg)}         &
\colhead{Half}                                          
}
\tablewidth{0pt}
\startdata
8591\dotfill \ldots \ldots & 2007 Sep 20, 05:26 & \phantom{0}45.4 & 03 32 28.20 & $-$27 48 06.9 & 72.7 & 1 \\
9593\dotfill \ldots \ldots & 2007 Sep 22, 20:34 & \phantom{0}46.4 & 03 32 28.20 & $-$27 48 06.9 & 72.7 & 1 \\
9718\dotfill \ldots \ldots & 2007 Oct 03, 13:56 & \phantom{0}49.4 & 03 32 28.61 & $-$27 48 07.4 & 62.0 & 1 \\
8593\dotfill \ldots \ldots & 2007 Oct 06, 02:04 & \phantom{0}49.5 & 03 32 28.61 & $-$27 48 07.4 & 62.0 & 1 \\
8597\dotfill \ldots \ldots & 2007 Oct 17, 07:07 & \phantom{0}59.3 & 03 32 29.25 & $-$27 48 10.4 & 44.2 & 1 \\
8595\dotfill \ldots \ldots & 2007 Oct 19, 14:16 & 115.4 & 03 32 29.35 & $-$27 48 11.2 & 41.2 & 1 \\
8592\dotfill \ldots \ldots & 2007 Oct 22, 12:14 & \phantom{0}86.6 & 03 32 29.62 & $-$27 48 13.8 & 32.4 & 1 \\
8596\dotfill \ldots \ldots & 2007 Oct 24, 13:20 & 115.1 & 03 32 29.62 & $-$27 48 13.8 & 32.4 & 1 \\
9575\dotfill \ldots \ldots & 2007 Oct 27, 05:43 & 108.7 & 03 32 29.62 & $-$27 48 13.8 & 32.4 & 1 \\
9578\dotfill \ldots \ldots & 2007 Oct 30, 22:35 & \phantom{0}38.6 & 03 32 29.84 & $-$27 48 16.7 & 24.2 & 1 \\
8594\dotfill \ldots \ldots & 2007 Nov 01, 11:51 & 141.4 & 03 32 29.84 & $-$27 48 16.7 & 24.2 & 1 \\
9596\dotfill \ldots \ldots & 2007 Nov 04, 04:11 & 111.9 & 03 32 29.95 & $-$27 48 18.5 & 19.8 & 1 \\
12043\dotfill\ldots\ldots & 2010 Mar 18, 01:39 & 129.6 & 03 32 28.78 & $-$27 48 52.1 & 252.2 & 1 \\
12123\dotfill\ldots\ldots & 2010 Mar 21, 08:08 &  \phantom{0}24.8 & 03 32 28.78 & $-$27 48 52.1 & 252.2 & 1 \\
12044\dotfill\ldots\ldots & 2010 Mar 23, 11:31 &  \phantom{0}99.5 & 03 32 28.55 & $-$27 48 51.9 & 246.2 & 1 \\
12128\dotfill\ldots\ldots & 2010 Mar 27, 13:08 &  \phantom{0}22.8 & 03 32 28.55 & $-$27 48 51.9 & 246.2 & 1 \\
12045\dotfill\ldots\ldots & 2010 Mar 28, 16:38 &  \phantom{0}99.7 & 03 32 28.32 & $-$27 48 51.4 & 240.2 & 1 \\
12129\dotfill\ldots\ldots & 2010 Apr 03, 15:21 &  \phantom{0}77.1 & 03 32 28.33 & $-$27 48 51.4 & 240.2 & 1 \\
12135\dotfill\ldots\ldots & 2010 Apr 06, 09:36 &  \phantom{0}62.5 & 03 32 28.01 & $-$27 48 50.2 & 231.7 & 1 \\
12046\dotfill\ldots\ldots & 2010 Apr 08, 08:17 &  \phantom{0}78.0 & 03 32 28.01 & $-$27 48 50.2 & 231.7 & 2 \\
12047\dotfill\ldots\ldots & 2010 Apr 12, 13:21 &  \phantom{0}10.1 & 03 32 27.80 & $-$27 48 48.9 & 225.2 & 2 \\
12137\dotfill\ldots\ldots & 2010 Apr 16, 08:53 &  \phantom{0}92.8 & 03 32 27.59 & $-$27 48 47.2 & 219.2 & 2 \\
12138\dotfill\ldots\ldots & 2010 Apr 18, 12:40 &  \phantom{0}38.5 & 03 32 27.59 & $-$27 48 47.3 & 219.2 & 2 \\
12055\dotfill\ldots\ldots & 2010 May 15, 17:15 &  \phantom{0}80.7 & 03 32 26.72 & $-$27 48 32.3 & 181.4 & 2 \\
12213\dotfill\ldots\ldots & 2010 May 17, 14:22 &  \phantom{0}61.3 & 03 32 26.69 & $-$27 48 31.1 & 178.9 & 2 \\
12048\dotfill\ldots\ldots & 2010 May 23, 07:09 & 138.1 & 03 32 26.64 & $-$27 48 27.6 & 171.9 & 2 \\
12049\dotfill\ldots\ldots & 2010 May 28, 18:58 &  \phantom{0}86.9 & 03 32 26.61 & $-$27 48 24.4 & 165.5 & 2 \\
12050\dotfill\ldots\ldots & 2010 Jun 03, 06:47 &  \phantom{0}29.7 & 03 32 26.61 & $-$27 48 21.7 & 160.2 & 2 \\
12222\dotfill\ldots\ldots & 2010 Jun 05, 02:47 &  \phantom{0}30.6 & 03 32 26.61 & $-$27 48 21.7 & 160.2 & 2 \\
12219\dotfill\ldots\ldots & 2010 Jun 06, 16:30 &  \phantom{0}33.7 & 03 32 26.61 & $-$27 48 21.7 & 160.2 & 2 \\
12051\dotfill\ldots\ldots & 2010 Jun 10, 11:30 &  \phantom{0}57.3 & 03 32 26.63 & $-$27 48 19.2 & 155.2 & 2 \\
12218\dotfill\ldots\ldots & 2010 Jun 11, 10:18 &  \phantom{0}88.0 & 03 32 26.63 & $-$27 48 19.2 & 155.2 & 2 \\
12223\dotfill\ldots\ldots & 2010 Jun 13, 00:57 & 100.7 & 03 32 26.63 & $-$27 48 19.2 & 155.2 & 2 \\
12052\dotfill\ldots\ldots & 2010 Jun 15, 16:02 & 110.4 & 03 32 26.70 & $-$27 48 14.5 & 145.7 & 2 \\
12220\dotfill\ldots\ldots & 2010 Jun 18, 12:55 &  \phantom{0}48.1 & 03 32 26.70 & $-$27 48 14.5 & 145.7 & 2 \\
12053\dotfill\ldots\ldots & 2010 Jul 05, 03:12 &  \phantom{0}68.1 & 03 32 27.02 & $-$27 48 06.0 & 127.0 & 2 \\
12054\dotfill\ldots\ldots & 2010 Jul 09, 11:35 &  \phantom{0}61.0 & 03 32 27.02 & $-$27 48 06.1 & 127.0 & 2 \\
12230\dotfill\ldots\ldots & 2010 Jul 11, 03:52 &  \phantom{0}33.8 & 03 32 27.02 & $-$27 48 06.0 & 127.0 & 2 \\
12231\dotfill\ldots\ldots & 2010 Jul 12, 03:22 &  \phantom{0}24.7 & 03 32 27.16 & $-$27 48 03.6 & 121.2 & 2 \\
12227\dotfill\ldots\ldots & 2010 Jul 14, 21:04 &  \phantom{0}54.3 & 03 32 27.16 & $-$27 48 03.7 & 121.2 & 2 \\
12233\dotfill\ldots\ldots & 2010 Jul 16, 10:25 &  \phantom{0}35.6 & 03 32 27.16 & $-$27 48 03.7 & 121.2 & 2 \\
12232\dotfill\ldots\ldots & 2010 Jul 18, 19:53 &  \phantom{0}32.9 & 03 32 27.16 & $-$27 48 03.7 & 121.2 & 2 \\
12234\dotfill\ldots\ldots & 2010 Jul 22, 19:58 &  \phantom{0}49.1 & 03 32 27.19 & $-$27 48 03.3 & 120.2 & 2 \\
\enddata

\tablecomments{All observations were done in VF mode. The last column refers to which half mosaic the exposure contributed to. The observations were split according to their total counts (see Table~\ref{tab:cts}). Half 1 has a total effective exposure time of 1483.7 ks, and half 2 has 1444.4 ks.}
\label{tab:obs}
\end{deluxetable*}

\begin{table}
\begin{center}
\caption{X-ray map counts}
\label{tab:cts}
\begin{tabular}{l  c  c  c  c  c  c }
\hline\hline
\vspace{.05cm}
Band & $N_{A}$ & $N_{B}$  & $N^{*}_{A}$ & $N^{*}_{B}$ & $\langle N \rangle$ pix$^{-1}$\\
\hline
soft & 25848 & 26799 & 9396 & 9917 & 0.18\\
hard & 72997 & 77424 & 32065 & 34116 & 0.51\\
full & 98845 & 104223 & 41462 & 44034 & 0.69\\
\hline
\end{tabular}
\end{center}
\textbf{Notes.}\\
Subscripts $A$ and $B$ denote each half map; asterisks denote masked maps.
\end{table}

\subsection{Self-calibration}
In order to measure fluctuations over large angular scales, we need to mosaic many individual exposures with smaller fields of view. This becomes especially challenging at NIR wavelengths because of the relatively high intensities of foreground emissions at these wavelengths. Sunlight scattered off of dust in the solar system, or ``Zodiacal light'', is a temporal foreground that will affect different frames exposed at different times of the year, and leads to large overall offsets between frames of the same or similar sky area. If these offsets are not properly modeled and removed, they will lead to a fictious anisotropy signal. Atmospheric airglow from the earth can also produce large offsets for HST observations that push its orbital limits and observe close to Earth's limb. The self-calibration algorithm \citep{Fixsen00, Arendt02, Arendt10} is a least-squares calibration algorithm that was explicitly designed to model and remove these offsets. So we utilize this code to generate our own mosaics, instead of the publicly available mosaics genarated by astrodrizzle \citep{Fruchter02, Astrodrizzle2012}.

As is described in \cite{Arendt02}, our data, $D^{i}$, are modeled as
\begin{equation}
\label{eq:selfcal}
D^{i} \approx G^{p}\,S^{\alpha} + F^{p} + F^{q}
\end{equation}
where $i$ indexes each of the pixels in the entire data set, $G^{p}$ is the gain of each detector pixel, and $S^{\alpha}$ is the sky intenisy at each sky position $\alpha$. $F^{p}$ and $F^{q}$ are offset terms per detector pixel (index $p$) and per frame (index $q$). If one can model their data in such a way, these last two interloper terms can be efficiently minimized or removed. The details of solving such an equation are nontrivial. We leave it to the reader to refer to \cite{Arendt02}, \cite{Arendt00} and \cite{Fixsen00} for the solution to eq.~(\ref{eq:selfcal}). It can be seen that for each additional frame included in the mosaics, the index $i$ increases by the number of pixels in each frame, which becomes exceedingly large for large area mosaics or small pixel scale frames (e.g. ACS). In these cases the self-calibration can take a considerable amount of time and computer memory to run.

\subsection{{\it Hubble} optical and NIR maps}
All of our HST data are publicly available, and were downloaded from the Barbara A. Mikulski Archive for Space Telescopes (MAST; located at https://archive.stsci.edu/hst/search.php). We assembled our own collection of calibrated, flat-fielded frames (FLT) from the MAST archive from ten different HST proposals \citep{Beckwith06, Giavalisco04, Grogin10, Koekemoer11, Windhorst11}. These data are collected in five different filters, collected with both the Advanced Camera for Surveys (ACS) and the infrared Wide Field Camera 3 (WFC3/IR). We did not necessarily include all of the frames from any one proposal. As the self-calibration is largely dependent on the dither pattern of the observations and requires significant pixel overlap \citep{Arendt00}, additional frames of the same pointing, from e.g. the Hubble Ultra Deep Field \citep{Beckwith06}, will not improve the self-calibration solution, but will increase the variance of the signal to noise of the mosaic (which will propogate into the power spectrum). We thus chose somewhat randomly from these observations, and included the full set of frames from the more wide and contiguous proposals \citep{Giavalisco04}. We also preferentially selected ACS frames that were exposed before the fourth servicing mission (SM4), as those data are not plagued with the horizontal bias striping that was introduced after the replacement of the readout box during SM4 \citep{Grogin10}. This striping is a spatially correlated noise and corrupts any spatial correlation measurements. We preformed simulations (MW15) to show that a mosaic produced with $\lesssim$~30\% of post-SM4 frames will be adequate to perform reliable angular correlation measurements.

All FLTs downloaded from MAST are reprocessed ``on-the-fly'', meaning they use the most recent calibration files to subtract bias and dark frames, and perform flat-field corrections. With our collection of FLTs in hand, we continue with the basic data reduction as prescribed by the Space Telescope Science Institue, using PyRAF version 2.1.1. First, all post-SM4 ACS frames are destriped; all FLTs are then charge transfer efficiency corrected. Cosmic rays are flagged using the CRCLEAN PyRAF module, and sub-arcsecond astrometric alignment against the publicly available CANDELS multidrizzle mosaics\footnote{http://candels.ucolick.org/data access/GOODS-S.html} is performed with the PyRAF module TWEAKREG.

Finally, we feed these reduced FLT frames to our self-calibration algorithm to generate mosaics. We use the same self-calibration model as in \cite{Arendt02}, another example of where HST data have been successfully self-calibrated. We de-weight bad pixels and cosmic arrays and iterate three times in order to find a self-calibration solution. Our input FLT frames are geometrically distorted with a pixel size of 0.$''$1354~$\times$~0.$''$120 for WFC3, and 0.$''$0498~$\times$~0.$''$0502 for ACS. We remove the distortion in the map making procedure and produce mosaics with a slightly larger geometrically square pixel size of 0.$''$14. As was previously mentioned, we perform jack-knife measurements, so each filter has two mosaics of the same sky area generated from different exposures. This way we can perform the cross-correlation of two maps and any uncorrelated noise will drop out of the measurements. For HST observations, each pointing has at least two exposures, so separating the data into two sets is straightfoward. We end up with two mosaics for each of the 0.6~\m~ (F606W), 0.7~\m~ (F775W), 0.85~\m~ (F850LP), 1.25~\m~ (F125W) and 1.6~\m~ (F160W) bands. The HST exposure times range between 180 and 1469 seconds; tile patterns for each filter can be seen in Fig.~1 of MW15.

\begin{figure*}
  \centering
  \includegraphics[scale=.55]{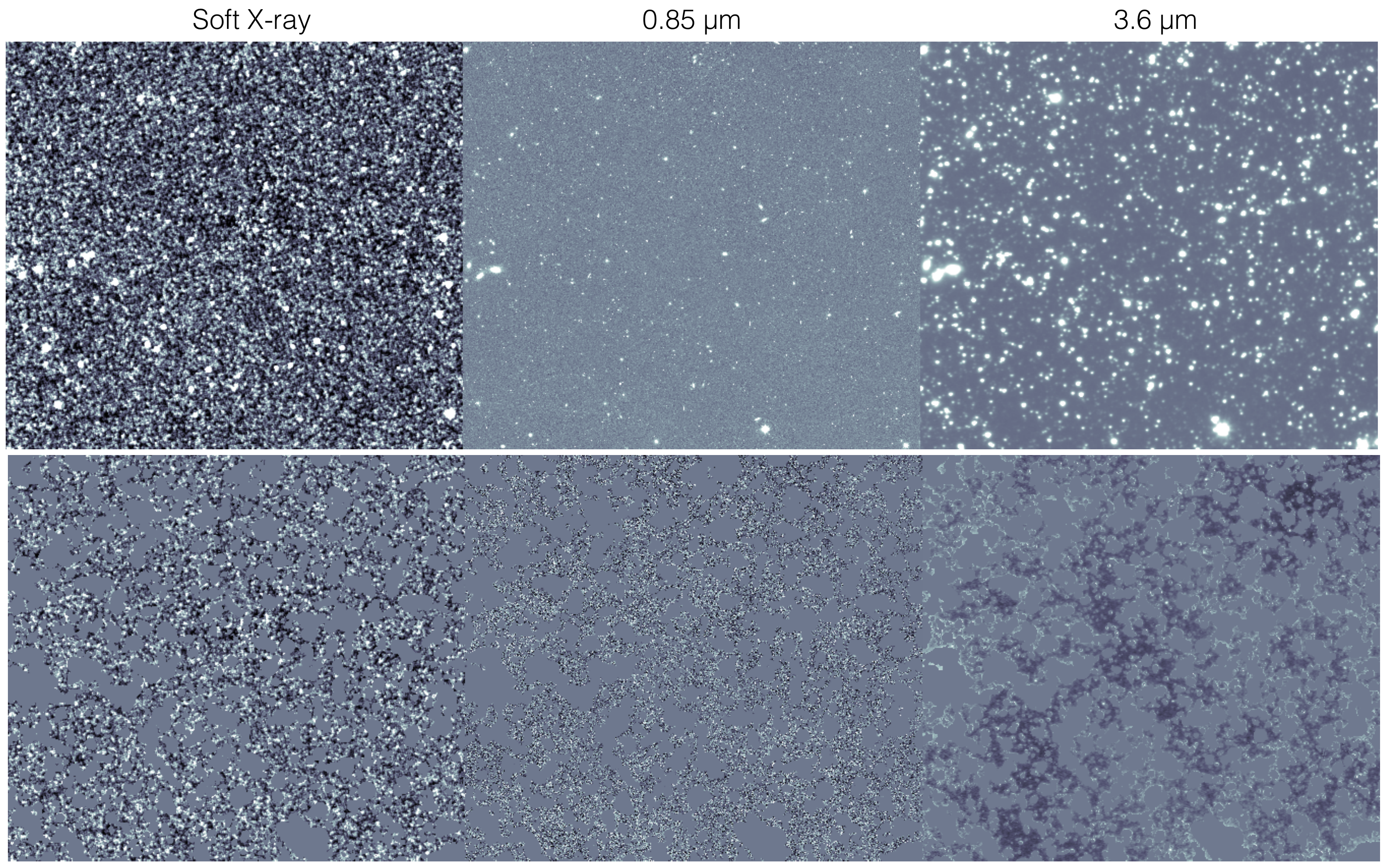}
  \label{fig:maps}
  \caption{A cropped section of the mosaics in each band, both unmasked (top) and median-subtracted masked (bottom). The arrays have been scaled arbitrarily for viewing purposes, with common mean values in the unmasked regions, and common standard deviations in the masked images.}
\end{figure*}

\subsection{{\it Spitzer} NIR maps}
{\it Spitzer}/IRAC Channel 1 (3.6~$\mu$m) and Channel 2 (4.5~$\mu$m) frames are also susceptible to the same foreground contaminats as the HST data. Therefore we use the self-calibration algorithm to generate the channel~1 and channel~2 mosaics. We assemble basic calibrated data (CBCD) from the Spitzer Heritage Archive. These data are generally of excellent quality and do not require significant reduction prior to mosaicing. All CBCDs are astrometrically aligned, and those from the cryogenic mission do not, in general, contain arfifacts. The only issue one needs to mitigate before performing the self-calibration relates to the warm mission CBCDs. The Spitzer Science Center (SSC) does not have lab darks to correct for different delay times in the warm mission, which introduces a ``stripiness'' into the frames. In addition, the sky dark frames that are subtracted from the data during the standard CBCD reduction pipeline usually over-subtracts the background, resulting in overall negative pixel values. The straightforward solution to both of these issues is to generate a median image from all the warm CBCDs in our set, and subtract it from each warm mission CBCD before running the self-calibration.

CBCDs are are collected from 6 different proposal IDs: 10076 (IGOODS; \citealt{Labbe15}), 60022 (SEDS; \citealt{Ashby13}), 61052 (SERVS; \citealt{Mauduit12}), 70145 (IUDF; \citealt{Labbe15}), 70204 (ERS; \citealt{Fazio11}), 80217 (S-CANDELS; \citealt{Ashby15}). We again do not necessarily use all the frames from each proposal; instead we limit those to a region just slightly larger than the CANDELS area (so the self-calibration solution converges more quickly). Each of the {\it Spitzer}/IRAC mosaics are generated from 8089 individual CBCD frames.

{\it Spitzer}/IRAC observations do not necessarily have multiple exposures for each pointing. To split the CBCDs into two halves, in order to make jack-knife maps, we simply sort all CBCDs in each filter according to their right ascension and take every other frame for each half. Apart from this sorting, and subtracting the warm median frame from warm mission CBCDs, we do not perform any additional reduction prior to self-calibrating.

\section{Power Spectra}
\subsection{Outline}
\label{sec:ps}
Standard fast fourier transform (FFT) techniques are used to measure the angular power spectrum on these small area maps. We're looking for correlations in the diffuse background light, so this requires generating a resolved source mask in all the bands, and quantifying the mode-mode coupling introduced by the mask. We also need to quantify the map-making and tiling patterns (transfer function), and correct for the finite resolution of the telescope (beam transfer function). Although these procedures are discussed in detail in other works \citep{Amblard11, Hivon02, Mitchell-Wynne15, Thacker13, Thacker15, Zemcov14}, we will discuss them here for clarity. 

\subsection{Resolved source mask}
Resolved sources are masked in the HST bands down to $\sim$ 27 $m_{\text{AB}}$; the mask generated for the work in MW15 is regridded to the larger pixel scale, which removes 47\% of the pixels. We use the public X-ray main catalog from \cite{Xue11} to generate an X-ray source mask, which in our field area corresponds to $\sim$ 200 sources. We mask within circular radii proportional to the logarithm of the source counts; this increases the masking percentage by $\sim$ 5\%. We do not include X-ray clusters in our source mask, as it would omit too many pixels. We mask IRAC sources down to $\sim$~24~$m_{\text{AB}}$ by convolving a 3$\sigma$ detected \texttt{SExtractor} mask with the IRAC beam. After taking the union of these three masks, we sigma clip each masked map at a 5$\sigma$ level, which removes an addtional $\sim$~1\% of the pixels, to obtain our final mask. These 1\% residual outliers are products of imperfect masking. In the end we remove $\sim$ 53\% of the pixels with the common source mask, which is applied to all bands and instruments. A cropped section of the source mask applied to the mosaics can be seen in the bottom panel of Fig~3.

When we apply the mask to our maps, it breaks some large scale modes into smaller scale modes, which needs to be corrected for. This correction is described in \cite{Hivon02}. We apply a mode-mode coupling correction with an \mll matrix (see Technical Supplement of \citealt{Cooray12}), which has been demonstrated with simulations \citep{Cooray12, Mitchell-Wynne15, Thacker15} to recover the true power spectrum of a masked map. We generate the \mll matrix with the same methods described in \cite{Cooray12}, using 100 simulations.

\subsection{$T(\ell)$ correction}
The next correction that is performed is the map-making transfer function, $T(\ell)$. For HST and {\it Spitzer}/IRAC we do this in a six step process:
\begin{enumerate}
  \item Generate simulated image with known power spectrum which spans the entire survey area.
  \item Fill each of the FLTs (for HST) or CBCDs (for {\it Spitzer}/IRAC) with the signal from the appropriate position on simulated map from the previous step.
  \item Add Gaussian noise proportional to the inverse square root exposure time to each FLT or CBCD.
  \item Add an overall offset equal to the median value of the science frame.
  \item Feed this set of FLTs or CBCDs to selfcal and generate a mosaic.
  \item Compare the input power spectrum to the output power spectrum to quantify $T(\ell)$.
\end{enumerate}
Step 4 ensures that any residual offsets in the mosaics are corrected for. This six-step procedure is done 100 times for each band; the error bars for our transfer functions are just the standard deviation at each $\ell$ bin. The $T(\ell$) for the HST mosaics can be seen in the Supplementary Section of MW15. The Spitzer $T(\ell)$ is plotted in Fig.~\ref{fig:tlbl}.

Since the X-ray maps aren't generated from frames plagued with offsets, we just need to check how the varying effective exposure time across the mosaic affects the power spectrum. This is done simply by generating 100 Gausian maps (constant $C_{\ell}$'s), multiplying each map with the effective exposure time map (shown in Fig.~\ref{fig:exposure}), and measuring how far from constancy the resulting power spectra are. Since the variance of the effective exposure times is small, this results in a transfer function that is roughly constant at all scales, as is shown in Fig.~\ref{fig:tlbl}. Error bars are again taken as the standard deviation at each bin.

\subsection{B($\ell$) correction}
\begin{figure}
  \includegraphics[scale=.66]{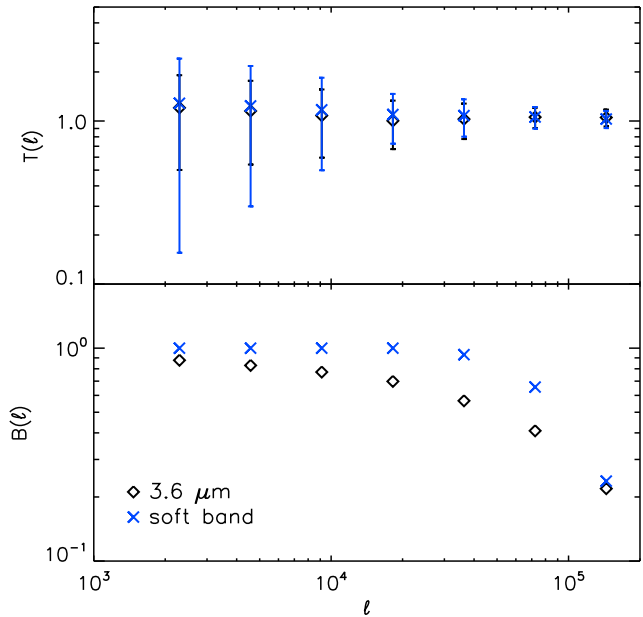}
  \caption{Map-making transfer functions (top) and beam transfer functions (bottom) for IRAC and the soft X-ray band. The different X-ray bands have slightly different effective exposure time maps, so the $T(\ell)$'s for each are unique but very similar. Beams for each X-ray band are roughly constant, with a 3 pixel-wide Gaussian PSF. Similar plots for the HST bands can be seen in Supplementary Figure~2 of MW15.}
    \label{fig:tlbl}
\end{figure}

Because of the finite resolutions of the telescopes used in this analysis, the power spectra will drop off at small angular scales. We correct this with a beam transfer function, $B(\ell)$. We measure the {\it Hubble}/WFC3 and ACS $B(\ell)$ using theoretical PSFs generated with {\sc Tiny Tim} \citep{Krist11}. The $B(\ell)$ is simply the FFT of these PSFs. The beam transfer functions measured in this way are consistent with those in MW15, and can be seen in the Supplementary Section of MW15. We chose to re-measure the $B(\ell)$'s using {\sc Tiny Tim} as a kind of sanity check that all was unchanged at the larger pixel scale used in this work. This larger pixel scale also made it more difficult to perform stacking, hence our use of the theoretical PSFs. For {\it Spitzer}/IRAC, we us the $B(\ell)$ that was measured in \cite{Cooray12}, and simply linearly interpolate those values at our $\ell$ bins. Those points are plotted in Fig.~\ref{fig:tlbl}

The point spread function (PSF) of \ch varies across the mosaic as a function of off axis angle. However, since we are only concentrating on a relatively small area and convolving by a relatively large beam, we do not need to quantify and correct for this PSF variation. The 50\% encircled energy average radius of the \ch PSF increases from $\sim$~0$''$ at the center of the image to $\sim$~3$''$ at an off axis angle of 8$'$ \footnote{\url{http://cxc.cfa.harvard.edu/proposer/POG/html/chap4.html\#fg:hrma\_ee\_offaxis\_hrci}}. The maximum off axis position of a \ch source in the CANDELS field is $\sim$~8.5$'$. Therefore we use the 3.6$''$ Gaussian convolution kernel as an approximation to the \ch beam. This approximation is further justified as we are not particularly interested in these small-scale fluctuations, which correspond to angular scales at $\ell\,> 3\,\times 10^{5}$. The X-ray $B(\ell)$ is then computed with a Gaussian of 3.6 arcseconds, which corresponds to the three-pixel Gaussian kernel that was used to convolve the X-ray maps. This curve can also be seen in Fig.!\ref{fig:tlbl}.

\subsection{Auto Spectra}
\begin{figure*}
  \centering
  \includegraphics[scale=.42]{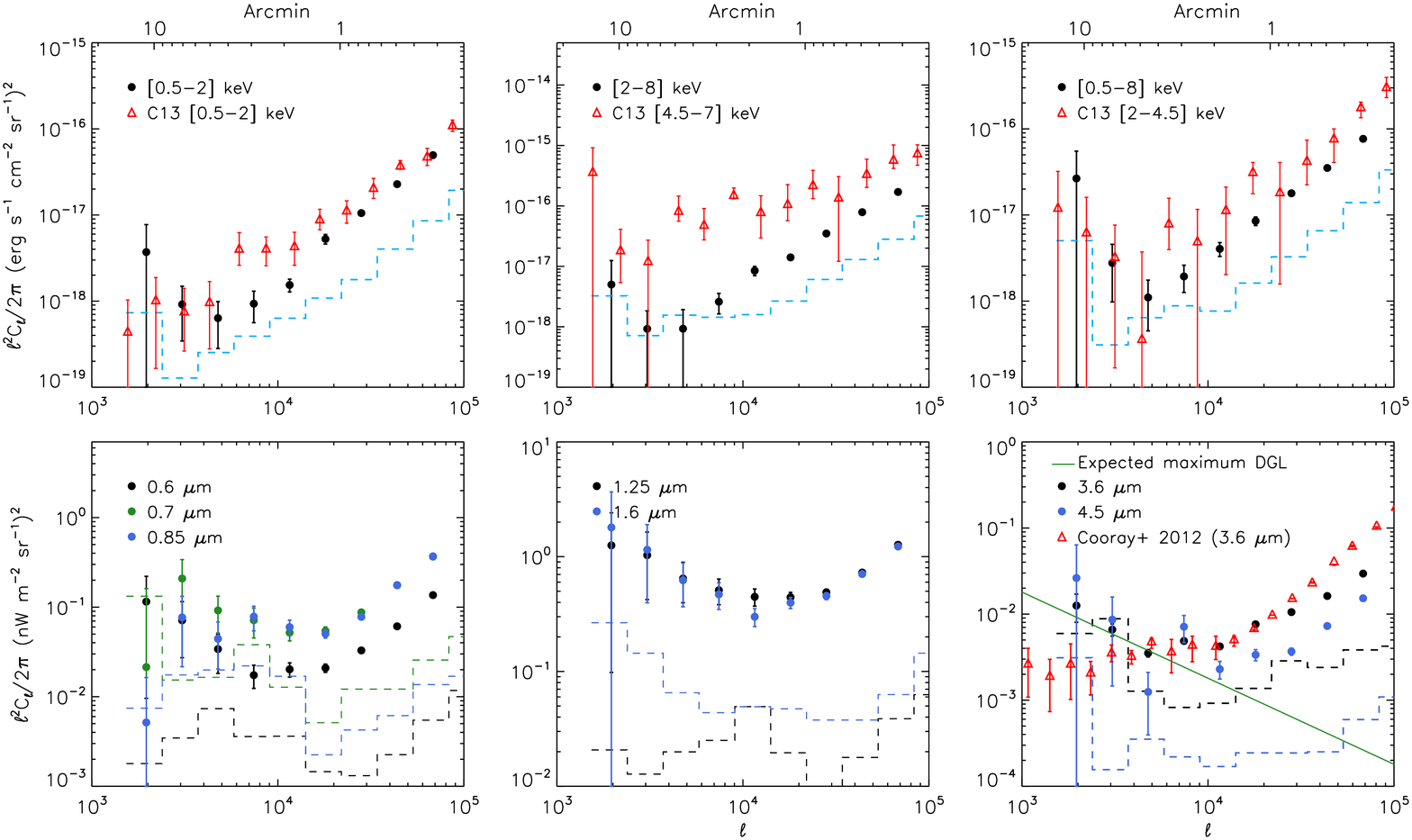}
  \caption{Auto power spectra in each band. Red triangles show measurements from previous works. Dashed lines show the noise power spectra (e.g. the auto spectra of (A-B)/2). The green line in the final plot shows the upper limit of the DGL contribution to the 3.6~\m~ auto-spectrum, modeled as $C_{\ell}^{\mathrm{DGL}} = A_{\mathrm{DGL}}^{\mathrm{3.6}}\,\ell^{-3}$.}
  \label{fig:auto}
\end{figure*}

For jack-knife maps $A$ and $B$ (see MW15 for details of jack-knife tests), we compute the auto-power spectrum as the cross-spectrum of $A\,\times\,B$, and quantify the noise-power spectrum as the auto spectrum of ($A-B$)/2. Taking the cross-spectrum will omit any uncorrelated noise. Cosmic variance is defined as 
\begin{equation}
    \delta C_\ell = \sqrt{ \frac{2}{f_{\mathrm{sky}}(2\ell+1)\Delta\ell}} (C_\ell^{\mathrm{auto}} +
    N_\ell),
\end{equation}
where $ N_\ell$ is the noise-power spectrum and $f_{\mathrm{sky}}$ is the fraction of the total sky unmasked in our maps. We compute our auto-spectra error bars by adding in quadrature the errors from the $T(\ell)$ and cosmic variance. The auto spectra for each of our 9 bands, including noise spectra, are plotted in Fig~\ref{fig:auto}.

C13 performed auto-correlation measurements between similar X-ray bands, and {\it Spitzer}/IRAC Channels 1 and 2. We compare our X-ray auto-spectra to theirs, and find a statistically significant disparity at angular scales of $\ell \sim 10^{4}$. The C13 auto-spectra have a bump at $\ell \sim 10^{4}$, where ours are dominated by the shot noise from unresolved AGNs at this scale. This discrepancy can likely be attributed to the difference in the two X-ray datasets. The CDF-S data used in this work are at a minimum 4 times deeper than the EGS/AEGIS data used in C13 \citep{Goulding12}. The broad X-ray band passes chosen in C13 are also slightly different than those that are used in this work.

\subsection{Cross-correlations: Measurements}
Cross-power spectra measurements are made by computing the cross-spectrum of ($A_{m}+B_{m})/2 \,\times\,(A_{n}+B_{n})/2$ for filters $m$ and $n$. The $\delta C_{\ell}$'s for a cross-correlation are computed as
\begin{equation*}
  \sqrt{ \frac{1}{f_{\mathrm{sky}}(2\ell+1)\Delta\ell} \left[
        (C_{\ell,m}^{\mathrm{auto}} + N_{\ell,m})(C_{\ell,n}^{\mathrm{auto}} + N_{\ell,n}) +
\left(C_\ell^{m \times n}\right)^2  \right]},
\end{equation*}
where $C_{\ell}^{m \times n}$ is the cross-correlation power spectrum between bands $m$ and $n$. Our total error budget is then the quadratic sum of the $\delta C_{\ell}$'s and the geometric mean of the transfer function between the two bands, e.g. $T_{\ell, mn} = \sqrt{T_{\ell,m}\,T_{\ell,n}}$. The beam transfer functions for the cross spectra are approximated in a similar way as $T_{\ell, mn}$.

\begin{figure*}[ht!]
  \centering
  \includegraphics[scale=.42]{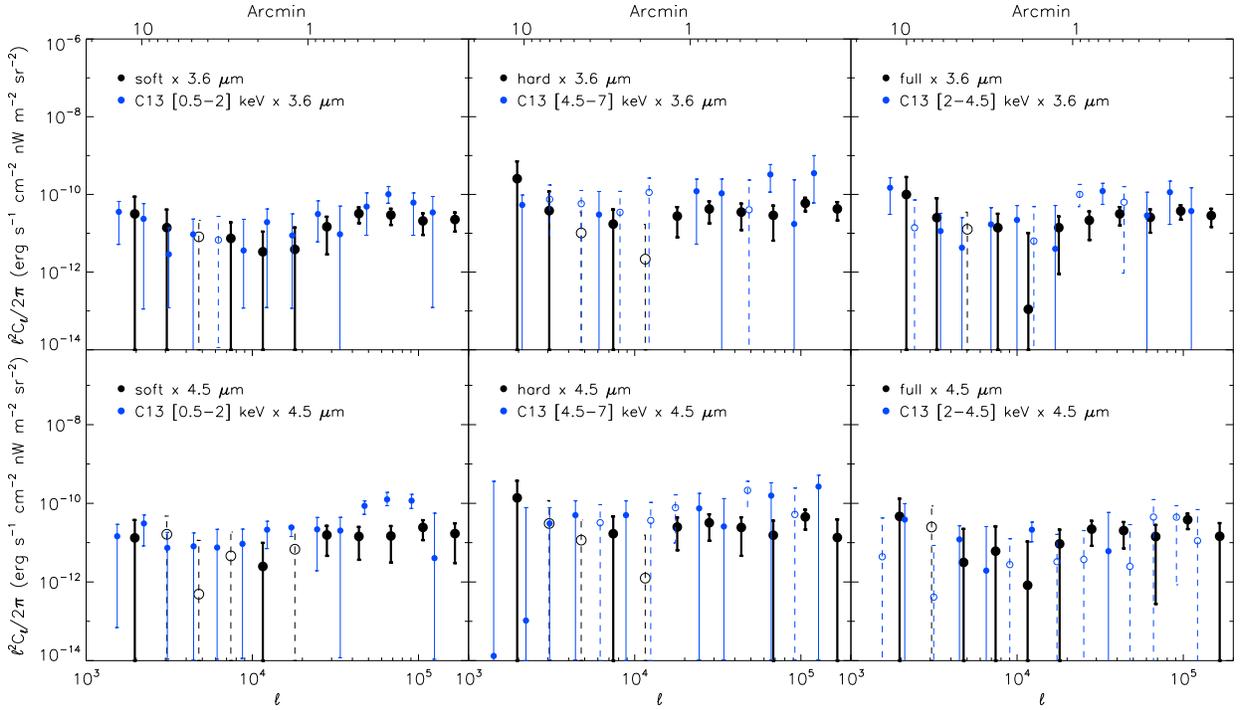}
  \caption{Raw cross-correlations of X-ray and IRAC 3.6~$\mu$m (top) and 4.5~$\mu$m (bottom). Open circles with dashed error bars denote negative values. Blue points are taken from Fig.~5 and 6 of C13. The spectra published in C13 are raw, so for consistency, here we compare our uncorrected spectra with theirs. There is good agreement between the two works.}
  \label{fig:raw}
\end{figure*}

\begin{figure*}[ht!]
  \centering
  \includegraphics[scale=.42]{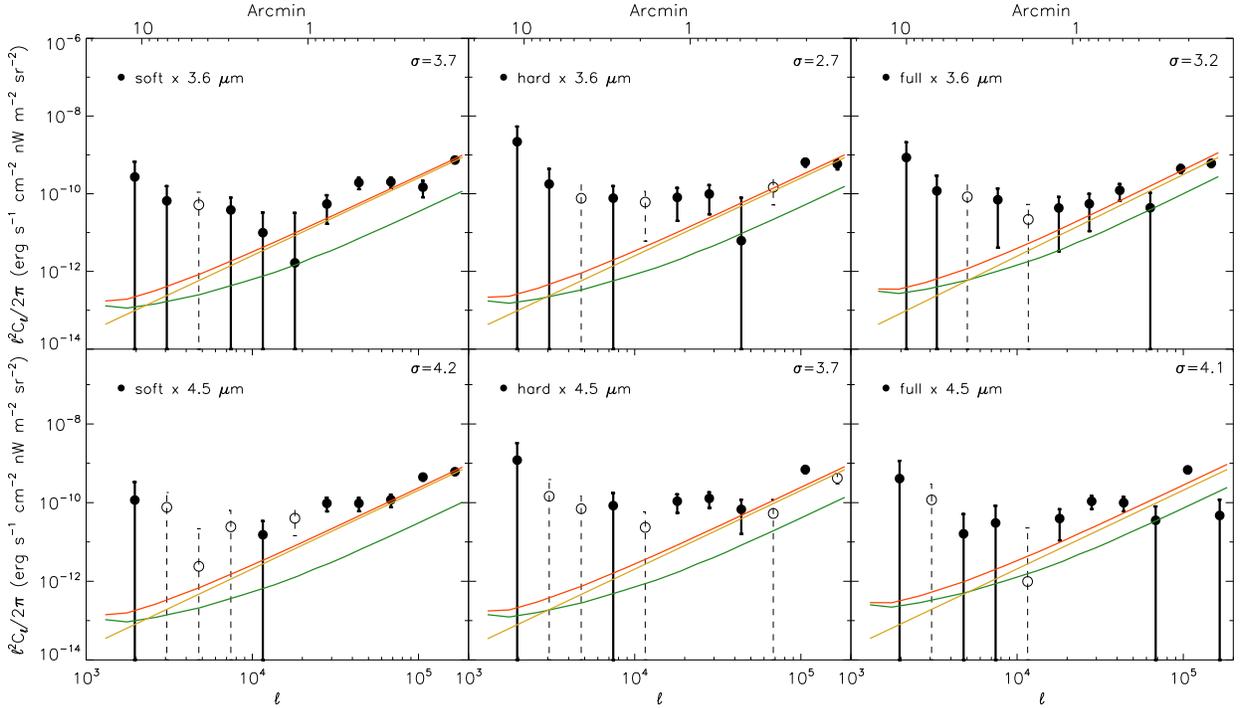}
  \caption{Corrected cross-correlations of X-ray and IRAC 3.6~$\mu$m (top) and 4.5~$\mu$m (bottom). Open circles with dashed error bars denote negative values. Blue points are taken from Fig.~5 and 6 of C13. The statistical significance of each cross-correlation, computed at angular scales $\gtrsim 20''$, is shown in the top-right of each plot. Theoretical models are shown as solid lines: the AGN component as a yellow line, unresolved galaxies as a green line, with a slight excess at low-$\ell$, and the total contribution as a red line.}
  \label{fig:ch1cross}
\end{figure*}

Since the measurements in C13 were presented with none of the corrections we consider here, we first compare our raw \cxb cross-correlations with IRAC 3.6 and 4.5~\m~ with theirs in Fig~\ref{fig:raw}. Our corrected \cxb cross-correlations can be seen in Figs~\ref{fig:ch1cross}. To be consistent with C13, we define the significance of the cross-correlation detections by concentrating only on angular scales larger than 20$''$ ($\ell \lesssim 6.5\,\times\,10^4$), however we compute these values from our corrected cross spectra, not the raw spectra. The maximum angular scale probed in our study is about 10$'$ and the significance is computed between $1.1\,\times\,10^3 \lesssim \ell \lesssim 6.5\,\times\,10^4$. We compute $p$-values from $\chi^2$ estimates of all the $C_{\ell}$'s within that range of angular scales. Our measured cross-correlations are generally in good agreement with those measured in C13. In that work they claimed a high significance with their soft band cross-correlations (3.8$\sigma$ at 3.6~\m~ and 5.6$\sigma$ for 4.5~\m), and lower significances in their cross-correlations with harder X-ray bands. We detect a similar result, with 3.7$\sigma$ and 2.7$\sigma$ significances for the soft and hard bands respectively at 3.6~\m, and 4.2$\sigma$ and 3.7$\sigma$ respectively at 4.5~\m (see Table~\ref{tab:sig}). In C13 they had a total of $\sim~1.3\times 10^5$ photons, where in our work we have $\sim~8\times 10^4$ (see Table~\ref{tab:cts}), which corresponds to a signal-to-noise ratio of about 1.3 times lower. The 3-4$\sigma$ significances that we measure here is more or less what one expects if we are seeing the same population as C13.

\begin{figure*}
  \centering
  \includegraphics[scale=.42]{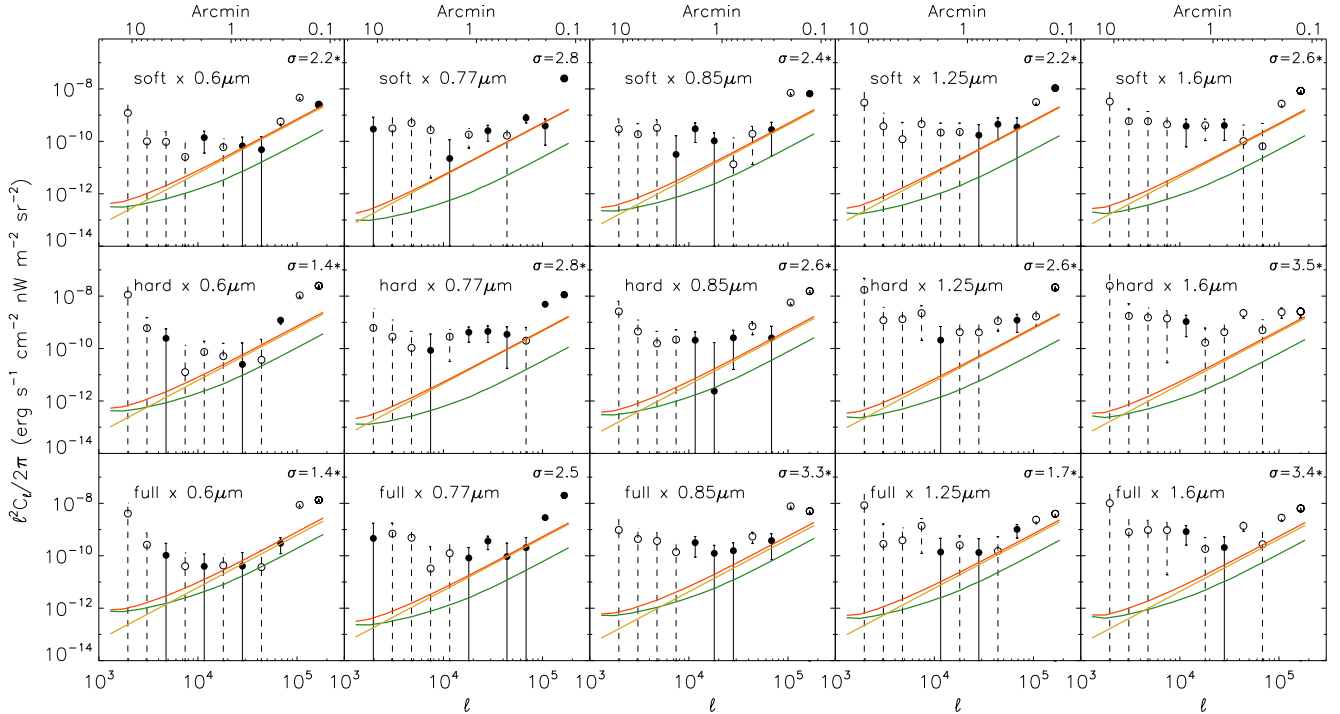}
  \caption{Cross-correlations of X-ray and optical/NIR HST maps. Open circles with dashed error bars denote negative correlations. The color schemes for the theoretical models curves are the same as in Fig.~\ref{fig:ch1cross}. The significance of each cross-correlations at angular scales $> 20''$ is shown in the top right of each plot. Sigma values with an asterisk denote an overall anticorrelation.}
  \label{fig:hstcross}
\end{figure*}

All 15 cross-correlations of the three X-ray bands with the five {\it Hubble} bands can be seen in Figure~\ref{fig:hstcross}. These are computed in the same manner as  discussed above for {\it Spitzer} fluctuations. The statistical significances for the five {\it Hubble} bands (as well as those from the 3.6 and 4.5~\m~ correlations) are shown in the top-right corners of their respective plots, and listed in Table~\ref{tab:sig}. Interestingly, we find the HST cross-correlations are largely anticorrelated, with the exception of two products at 0.775~\m, with anticorrelation significances ranging from 1.4 - 3.5$\sigma$. The two positive correlations with 0.775~\m~ have significances of 2.5 and 2.8$\sigma$. Statistically, all these (anti) cross-correlations are less significant, but we note that 10 out of the 15 cross-correlations below 1.6 $\mu$m are negative.
The presence of anticorrelations suggest that the fluctuation regions that are bright in the optical are faint in the X-rays, for example. One possibility for this is the Galactic absorption of soft X-rays, as discussed in \cite{wang95} using ROSAT and IRAS, and \citep{Snowden00}. While the extragalactic soft X-ray background is absorbed by the Galactic clouds, the same clouds (and associated cirrus) contributes positively to infrared fluctuations through DGL, Galactic dust-scattered interstellar light. While our detections of the anti cross-correlations are statistically less significant, the general behavior is consistent with suggestions in the literature that optical and IR background fluctuations can be impacted by Galactic signals such as DGL (MW15, \citealt{Yue16}).

\begin{figure*}
  \centering
  \includegraphics[scale=.55]{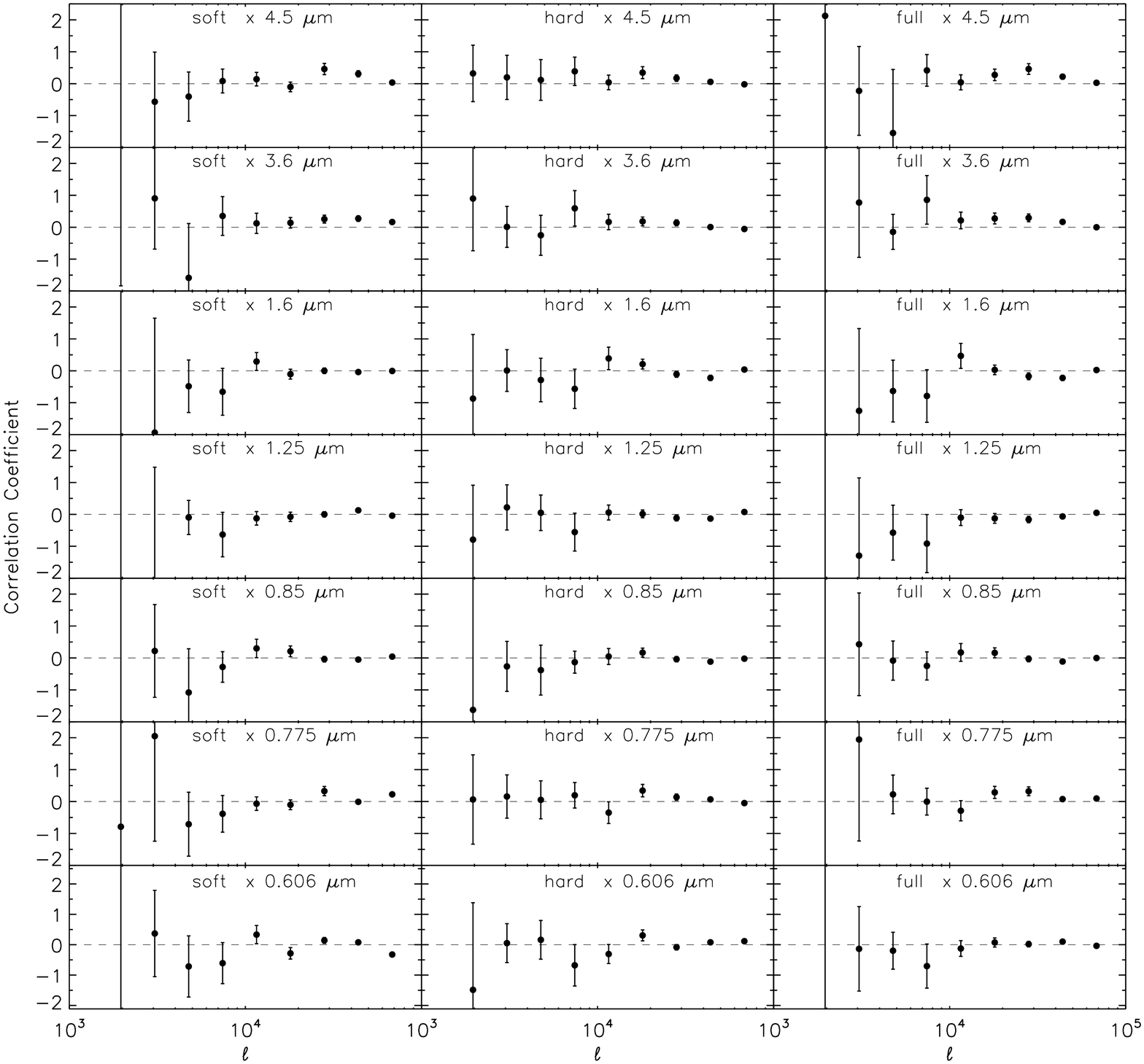}
  \caption{Correlation coefficients of X-ray and optical/NIR HST cross-correlations. All the points are statistically between -1 and 1.}
  \label{fig:coefs}
\end{figure*}

In addition to these cross-correlations, we show the correlation coefficient for each of the 21 cross-correlations. For bands $n$ and $m$, this coefficient is defined as $C_{\ell}^{n\times m}/\sqrt{C_{\ell,n}^{\mathrm{auto}}\,C_{\ell,m}^{\mathrm{auto}}}$. These are shown in Fig~\ref{fig:coefs}. The correlation coefficient can be interpreted as the fraction of total emission that is common between the two populations at filters $m$ and $n$, which means the physical source of emission is the same between the two bands, or that separate emitters at band $m$ and $n$ (e.g. IR and X-ray) are spatially separated by an angle less than the beam size. 

C13 and \cite{Helgason14} used their measured values of the correlation coefficient at large angular scales of the soft X-ray band and the 4.5~$\mu$m band to infer a lower limit fraction of BH emission in the CIB at that 4.5~$\mu$m band. Their values are quoted without errors that $\simeq$~15\%-25\% of the emission in that IR channel is a product of BH emission. As can be seen in Fig~\ref{fig:coefs}, the correlation coefficients we measure come with sufficiently large errors that we will not make any definitive statements about what fraction of the CIB emission comes from low-$z$ or high-$z$ sources and/or BHs. For example, the correlation coefficient we measure at large angular scales for the soft band and 4.5~$\mu$m ranges from $-0.56 \pm 1.55$ to $2.62 \pm 6.86$ with an average value of $0.32 \pm 0.66$.

\begin{table*}
\begin{center}
\caption{Statistical significance ($\sigma$) for $\ell \gtrsim 20''$}
\label{tab:sig}
\begin{tabular}{l | c | c | c | c | c | c | c}
\hline\hline
& 0.6 \um & 0.7\um & 0.85 \um & 1.25 \um & 1.60 \um & 3.6 \um & 4.5 \um\\ 
\hline
soft &  2.2$^{*}$ &  2.8      &  2.4$^{*}$ & 2.2$^{*}$ & 2.6$^{*}$ &  3.7 &  4.2\\ 
hard &  1.4$^{*}$ &  2.8$^{*}$ &  2.6$^{*}$ &  2.6$^{*}$ &  3.5$^{*}$ &  2.7 &  3.7\\ 
full &  1.4$^{*}$ &  2.5      &  3.3$^{*}$ &  1.7$^{*}$ &  3.4$^{*}$ &  3.2 &  4.1\\ 
\hline

\end{tabular}\\
Values with an asterisk denote anticorrelations.  
\end{center}
\end{table*}

\subsection{Cross-correlations: Model Predictions}
We show in Fig.~\ref{fig:hstcross} the expected contribution of unresolved AGN and galaxies (from their X-ray binary populations) based on the population models of \cite{Helgason14}. In this study, source emission was modeled separately in the near-IR and \mbox{X-rays} for both galaxies and AGN at $z<6$, to derive the unresolved background for a common NIR/X-ray masking threshold. The angular cross power spectra were calculated using a halo model and validated by an independent semi-analytical model mapped onto N-body simulations \citep{Henriques12}. For this paper, we extend these models to include optical wavelengths as well as hard X-ray energies using the same methodology as described in \cite{Helgason14}.

The models are displayed as solid lines alongside the measurements in Figures~\ref{fig:ch1cross} and \ref{fig:hstcross}. Whereas the galaxy contribution shows some increased power from clustering toward large angular scales, the AGN contribution is almost purely shot noise dominated. This is because many AGN that are relatively X-ray bright will still remain unresolved if they are optically dim. However, the two contributions are similar in amplitude because the abundance of the two source classes is comparable at the depth of the ECDFS. The models are somewhat sensitive to the extent to which the IR mask removes sources that remain undetected in X-rays.

As shown in Figs 6. and 7 existing models for the X-ray background, with simple populations involving faint galaxies and AGNs, are adequate to explain the measured cross-correlations with {\it Spitzer}. These models, however, do not predict the anti-correlation behavior seen at wavelengths below 1.6 $\mu$m since they do not involve Galactic absorption of X-rays. We do not pursue such detailed modeling here as we do not have adequate statistics to constrain additional model parameters. In the future with wide area surveys both in the X-rays (e.g., with Athena) and optical/infrared (such as with Euclid and WFIRST) the study can be expanded to 10-100 deg$^2$ areas for sufficient statistics for studies we have alluded here. It should also be noted that the evolution of X-ray binaries may be complex \citep{Lehmer14}, which is not considered in the modeling.

\section{Discussion}
\label{sec:discussion}
A previous study focusing on the {\it Spitzer}/Extended Groth Strip (EGS; \citealt{Davis07}) found $ \sim 3.8\sigma$ and $5.6\sigma$ cross-correlation detections at large angular scales between unresolved fluctuations at {\it Spitzer}/IRAC 3.6 and 4.5~$\mu$m respectively, and the {\it Chandra} soft X-ray band of 0.5 to 2 keV  (C13). This cross-correlation has been explained as due to a population of primordial DCBHs at $z > 12$ \citep{Yue13b}. If DCBHs explain all of the IRAC fluctuations at 3.6~$\mu$m, due to the Lyman-break cut-off redshifted to infrared wavelengths today, we do not expect to see any fluctuations generated by DCBHs at wavelengths less than 1.6~$\mu$m. Despite this general expectation, the recent study by MW15 found a significant detection of the intensity fluctuations from optical to infrared bands (0.6 to 1.6~$\mu$m) with {\it Hubble}/ACS and WFC3 imaging data in the GOODS and CANDELS surveys. These fluctuations, as well as previous detections of fluctuations at 3.6~$\mu$m, have been explained as mostly due to IHL (Cooray et al. 2012), with some evidence for fluctuations generated by galaxies present during reionization (MW15). Since IHL is the signal from tidally stripped stars, it will not have any X-ray component directly associated with it.

In the present study we return to the claimed detection between X-rays and {\it Spitzer}/IRAC. We extend the analysis to consider multi-wavelength optical and infrared data between 0.6 and 1.6 $\mu$m, in addition to 3.6 and 4.5~$\mu$m data in the CDF-S. Here we have complied and produced sky maps in the CDF-S in 10 passbands, from 3 different space telescopes, corresponding to 3 bands at X-ray wavelengths and 7 bands between 0.6 to 4.5~\m~with both {\it Hubble} and {\it Spitzer}. After masking the resolved sources to isolate the background light with a common mask, which itself involves the product of individual source-detection masks in each of the wavelengths, we performed cross-correlations of the X-ray maps with the optical and NIR data. 

The X-ray maps are sensitive to emission from black hole accretion and hot ionized gas, such as that found in galaxy clusters.  The optical and infrared maps trace the faint galaxies throughout the cosmic history. The redshift dependence of the galaxy poupulation is captured by the Lyman-break signal that is moving across the bands from 0.5 to 1.6 $\mu$m as a function of the redshift. The optical and infrared fluctuations also contain a signature of DGL, arising from dust-scattered light in our Galaxy. We found a 3.7-4.2$\sigma$ correlation between the soft X-ray band with both 3.6 and 4.5~\m, confirming the general result of C13. As the measurement is in an independent field, and utilizes independent analysis techniques and methods, we can generally confirm that there is indeed a significant cross-correlation between unresolved infrared fluctuations at 3.6, 4.5~$\mu$m and soft X-rays. Similar to C13, we also find a lower statistical significance in cross-correlation between the hard X-ray bands and IRAC.

Extending the analysis in C13, we also present results related to 0.6 and 1.6 $\mu$m fluctuations and find that neither soft nor hard X-ray bands are correlated with the optical or infrared fluctuations at a statistically significant level ($\gtrsim$~3.5$\sigma$). Those cross-correlations are mostly anti-correlated. \cite{wang95} and \cite{Snowden00} found shadows in the CXB introduced by the galaxy. We can then attribute, at least in part, these anti-correlations between HST and \ch to \mbox{X-ray} absorption by dust in the Milky Way (diffuse Galactic light; DGL), which is present in the NIR HST maps. However if this is the mechanism responsible for our measurements, it would also affect the the cross-correlation between the soft band and 3.6-4.5~$\mu$m, which would imply our measured cross-correlation significances at those wavelengths are lower limits. Furthermore it is unlikely this interpretation can be extended to the hard X-ray band since the galaxy is largely transparent at those wavelengths. The Galactic signal is present between 1.6 and 3.6 $\mu$m, complicating an easy interpretation of the auto and cross power spectra. Due to the lack self-calibrated maps at wavelengts between 1.6 and 3.6~$\mu$m, we do not have a complete understanding of the exact spectral energy distribution of the component that is correlated with the X-ray background.

\cite{Giallongo15} claimed a population of AGNs at $z > 4$ could be part of the population responsible for reionization. From their work, 7 of those sources are outside our cropped field of view and the remaining 15 sources are behind our source mask. So those purported signals do not contribute to the measurements reported here. \cite{Treister13}, \cite{Weigel15} and \cite{Cappelluti16} all find weak or no evidence of a significant population of AGNs at $z \gtrsim 4-5$, which would be consistent with our results assuming a high-redshift population of X-ray sources exist. In that case, the X-ray cross-correlation below 1.6~$\mu$m should disappear, due to the Lyman-break cut-off. In addition, \citep{Kashlinsky07} found no correlation with IRAC and faint ACS galaxies, which also aligns with our measurements. If there was a significant cross-correlation between those bands, it would follow that they would both be correlated, to some degree, with the same X-ray maps. The cross-correlation between the CIBER wavelengths and {\it Spitzer}/IRAC \citep{Zemcov14} can be wholly attributed to DGL \cite{Yue16} and thus does not add contradiction to this interpretation or previous findings.

The signal-to-noise of our cross-correlation measurements are ultimately limited by the number of X-ray photons, so larger area fields will improve the cross-correlation statistics. However large area, self-calibrated HST maps, generated from e.g. the 1.7 deg$^2$ F814W COSMOS observations (\citealt{Koekemoer07, Scoville07}), have proven difficult if not impossible to generate with the current number of exposures. Although the weak correlation coefficient between HST and {\it Spitzer}/IRAC (compounded with a weak X-ray/HST cross-correlation) does suggest the background sources of {\it Spitzer}/IRAC and HST are dissimilar, it is plausible that our HST/X-ray cross-correlations are just not sensitive to a signal that is actually there, because of the limited number of X-ray photons. It is also plausible that the X-ray background sources reside at $z>13$, which the longest HST band (1.6~\m) would not be sensitive to.

Our result is consistent with the existing framework related to our understanding of the nature and origin of optical and infrared intensity fluctuations, but with some modifications. Part of the component that is correlated between X-rays and 3.6, 4.5~$\mu$m could be in the form of DCBHs or some other source population. For example, apart from DCBHs, the soft X-rays can also come from hot gas associated with either star-forming regions or halos around galaxies, both of which may be expected to correlate with 3.6 and 4.5~\m. Due to the presence of optical fluctuations, we also require a signal from low redshifts and, in current models, such an origin involves IHL or tidally-stripped diffuse stars that populate the extended dark matter halos of galaxies. A combination of this IHL component and the component that is correlated with X-rays, such as DCBHs, likely contribute to fluctuations at 3.6 and 4.5~$\mu$m. Despite significant progress and overall improvement in the statistical accuracy in recent years, the cross-correlation between {\it Spitzer}/IRAC and X-rays is at the level of 4$\sigma$ significance and we are not able to model-fit to establish accurately the relative fraction between IHL and DCBH amplitudes at 3.6 and 4.5~$\mu$m. The lack of a significant cross-correlation between 1.6 and 3.6~$\mu$m could be interpreted as DCBHs primarily dominating the fluctuations at 3.6 and 4.5~$\mu$m, while IHL is dominating the fluctuations at 1.6~$\mu$m. Given the overall statistical uncertainties in the cross-correlation between 1.6 and 3.6~$\mu$m, even a scenario in which 100\% of the fluctuations are from IHL in 3.6 and 4.5~$\mu$m cannot be ruled out by the present measurements. To separate IHL from DCBHs, we need to improve not only the sensitivity to large-scale fluctuations in infrared and X-ray data, but also improve methods to separate reliably the DGL signal in the fluctuation power spectra. Naturally, additional deeper and wider fields in X-rays, and in the optical between 0.5 and 4 $\mu$m, will be one way to improve the current situation. In the future, appropriate data will likely come from survey telescopes such as {\it Athena} for X-rays and Euclid/WFIRST for optical and infrared.

\section{Summary}
In this paper we have measured the cross-correlations between the CXB and CIB and found a correlation between the soft X-ray background and the {\it Spitzer}/IRAC 3.6 and 4.5~\m~ backgrounds only, which is consistent with the previous findings of C13. Extending the results of C13, we find X-ray cross-correlations between the shorter wavelengths, ranging from 1.6~\m~ down to 0.6~\m, are statistically insignificant or anticorrelated. All correlations with the hard X-ray band were also found to be less significant or anticorrelated. This result implies that a significant portion of the CIB at 3.6 and 4.5~\m~could be a product of DCBHs; however it does not simultaneously rule out a partial or full contribution from IHL in the 3.6 and 4.5~\m~intensity fluctuations. Furthermore if the DCBHs do reside at $z > 13$, we would expect no correlations in bands shortward of 1.6~\m~with X-ray, which is consistent with our measurements. This lack of correlations in the shorter bands may be evidence that the sources responsible for the 3.6 and 4.5~\m~ backgrounds are fundamentally different from those at 1.6~\m~ and below.

Given the overall low significance of the detections in the near-IR and lack of detections in the optical, it is hard to draw strong conclusions from this study. We recommend further studies with wider X-ray fields of areas greater than 4 sq. degrees, since these studies benefit from the wide area coverage probing the linear clustering and not the narrow, deep fields more sensitive to the shot-noise. We recommend Chandra conduct a sufficient survey for the purposes of a cross-correlation study in a field overlapping with sufficient ancillary data in the near-infrared.

\vspace{.2cm}
We thank K\'{a}ri Helgason for performing the theoretical modeling of the cross-spectra, and for useful comments and discussions pertaining to this manuscript. We also thank the anonymous referee for thoughtful insights and useful suggestions to improve this manuscript. This work is based on observations taken by the CANDELS Multi-Cycle Treasury Program with the NASA/ESA HST, which is operated by the Association of Universities for Research in Astronomy, Inc., under NASA contract NAS5-26555. K.M.W. acknowledges support from HST-AR-13886, NASA grants NNX16AF39G, NNX16AJ69G and NNX15AQ06A, and GAANN P200A150121. A.C. acknowledges support from NSF CAREER AST-06455427, AST-1310310, and STScI Archival Research program GO HST-AR-13886.001-A. BL and WNB thank the Chandra \mbox{X-ray} Center for grants GO4-15130A and AR3-14015X. YQX acknowledges support of the Thousand Young Talents program (KJ2030220004), the 973 Program (2015CB857004), the USTC startup funding (ZC9850290195), the National Natural Science Foundation of China (NSFC-11473026, 11421303), the Strategic Priority Research Program ``The Emergence of Cosmological Structures'' of the Chinese Academy of Sciences (XDB09000000), and the Fundamental Research Funds for the Central Universities (WK3440000001).

\bibliography{xray}

\appendix

\section{Null Tests} 
\begin{figure*}
  \centering
  \includegraphics[scale=.84]{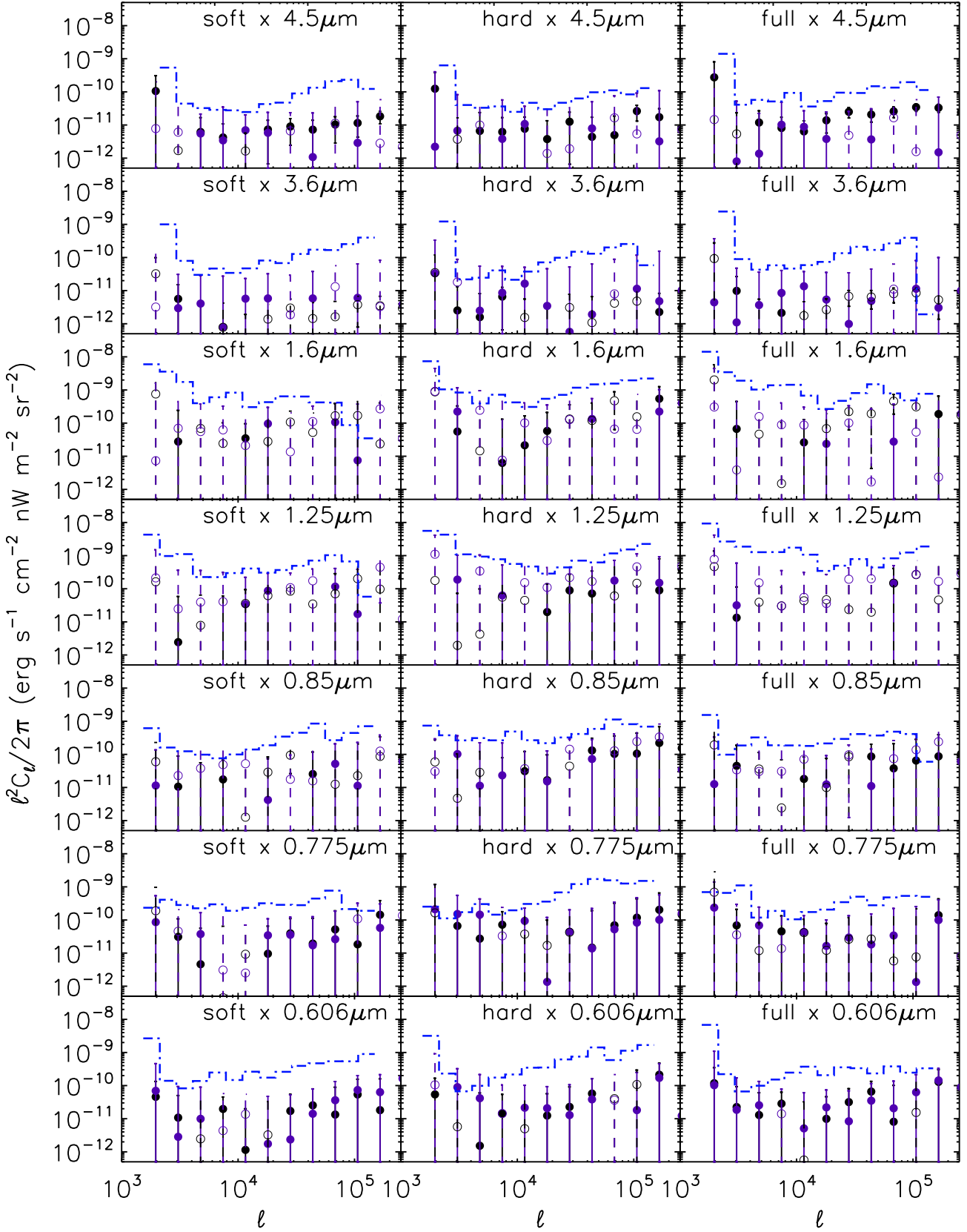}
  \caption{Null and signal map cross-correlations of X-ray and optical/NIR maps. Open circles with dashed error bars denote negative correlations. The black points correspond to the null (A-B)/2 NIR or optical map cross-correlated with the X-ray signal map. The purple points correspond to the null X-ray map cross-correlated with the signal (A+B)/2 NIR or optical map. The dot-dashed blue line is the absolute value of the signal-signal cross-correlations presented throughout this work.}
  \label{fig:nulls}
\end{figure*}

In this section we briefly address a kind of ``safety test'', in that we cross-correlate NIR/optical null maps with X-ray signal maps and vice versa. For jack-knfe maps $A_{1}$ and $B_{1}$ for band 1, and $A_{2}$ and $B_{2}$ for band 2, we recompute the cross-correlations twice for each combination of null and sky map signals: $(A1-B1)/2$~$\times$~$(A2+B2)/2$ and $(A1+B1)/2$~$\times$~$(A2-B2)/2$. We do this for each cross-correlation considered in this work, shown in Fig~\ref{fig:nulls}. We show in that plot the null tests, and for comparison the sky map cross-correlations presented throughout this work. The sky map raw cross spectra are generally an order of magnitude or more above the null test cross-correlations, not dissimilar to what was found in C13. This is a good test that spurious instrumental features are not dominating our sky map cross-correlations. Note that these null tests are encapsulated in our error bars throughout this work, as they do have contributions from the noise power spectra, discussed in \S 3.5.

\end{document}